\documentclass[
  reprint,
  superscriptaddress,
  amsmath,amssymb,
  aps,prl
]{revtex4-2}

\usepackage{graphicx}
\usepackage{dcolumn}
\usepackage{bm}
\usepackage{xcolor}
\usepackage{ulem}

\begin{document}

\title{Dehydration-Driven Ion Aggregation and the Onset of Gelation in ZnCl$_2$ Solution}

\author{Alexei V. Tkachenko}
\email{oleksiyt@bnl.gov}
\affiliation{Center for Functional Nanomaterials, Brookhaven National Laboratory, Upton, NY 11973, USA}
\author{Chuntian Cao}
\affiliation{Artificial Intelligence Department, Brookhaven National Laboratory, Upton, NY 11973, USA}
\author{Amy C. Marschilok}
\affiliation{Interdisciplinary Science Department, Brookhaven National Laboratory, Upton, NY 11973, USA}
\affiliation{Institute of Energy: Sustainability, Environment and Equity, Departments of Materials Science and Chemical Engineering, Stony Brook University, Stony Brook, New York 11794, USA}
\author{Deyu Lu}
\email{dlu@bnl.gov}
\affiliation{Center for Functional Nanomaterials, Brookhaven National Laboratory, Upton, NY 11973, USA}

\begin{abstract}
A minimal model of ionic aggregation in concentrated ZnCl$_2$ is developed, guided by molecular dynamics simulations with a machine-learned potential. It explicitly incorporates solvent-site depletion, correlated chloride binding, and allows for loops within Zn-Cl clusters. Dehydration is shown to drive ion binding through two sharp transitions set by the Zn coordination number $Z$: a crossover at $Z=2$ from isolated ions to Cl-bridged clusters, and gelation near $Z\approx 3$. The model agrees quantitatively with MD results, and the critical exponent of the cluster-size distribution matches percolation theory. 
\end{abstract}
\maketitle
Highly concentrated “water-in-salt’’ (WIS) electrolytes exhibit a physicochemical behavior that lies well outside the scope of classical dilute-ion descriptions. Among them, ZnCl$_2$ stands out as one of the most soluble inorganic salts containing a multivalent cation \cite{JI202199}. Its exceptional solubility and anomalous properties are widely attributed to the formation of stable multi-ion complexes such as ZnCl$_4^{2-}$. These complexes also enable highly concentrated ZnCl$_2$ solutions to dissolve cellulose, a property exploited since the 19$^{\text{th}}$ century \cite{JI202199}. Beyond its importance as a model WIS system, ZnCl$_2$ is technologically relevant for next-generation zinc-ion batteries, which were found to suppress the water splitting reactions, help regulate the morphologies of Zn plating, and improve the Coulombic efficiency of Zn anodes~\cite{wang2018highly}.

Recent atomic simulations~\cite{Cao_PRXEnergy} suggest that Zn ion aggregates can form in the WIS regime with a strong impact on ion transport, yet the microscopic structure and connectivity of multi-ion aggregates in the WIS regime remain poorly understood. This behavior is qualitatively consistent with earlier theoretical models \cite{jcp2020, McEldrew_2021,PRXEnergy2023} that draw analogies between ionic aggregation and classical Flory–Stockmayer (FS) \cite{Flory1941,Stockmayer1944,Tanaka1994} theories to describe super-concentrated ionic solutions. In those models, ions and solvent molecules are represented as patchy particles characterized by a maximum number of bonds (“patches”), and  bond formation is taken to be independent among patches and determined solely by the availability of unbound sites, producing loopless Cayley-tree clusters with fixed degree distributions. 

In the present paper, we combine molecular dynamics (MD) simulations based on a machine-learning interatomic potential ~\cite{Cao_PRXEnergy} with an analytic model to uncover the mechanism behind ionic aggregation in ZnCl$_2$ WIS electrolytes. Our minimal model is conceptually related to the FS-based picture \cite{Flory1941,Stockmayer1944,Tanaka1994,jcp2020, McEldrew_2021,PRXEnergy2023}, but incorporates several essential revisions motivated directly by our MD results. \textit{First}, we lift the Cayley-tree constraint and allow the formation of loops, a key modification, since the presence of loops places our system in a different universality class from the loopless  FS framework within the broader landscape of percolation problems. \textit{Second}, we relax the assumption of independent bond formation, specifically for chloride ligands, in order to capture the substantial free-energy penalty associated with forming a Cl-mediated Zn--Cl--Zn bridge. This correlated binding produces qualitatively distinct behavior, including the sharp coordination-driven crossover that we identify as the \textit{aggregation threshold}. \textit{Third}, guided by the known structure of Zn$^{2+}$ and Cl$^{-}$ hydration shells and supported by MD, we allow the effective topological valence of each species to differ between ion-ion and ion-water interactions. \textit{Finally}, we treat the availability of hydrogen and oxygen sites of the water molecules as independent degrees of freedom, consistent with classical FS theory,  but differently from the generic model of McEldrew \textit{et al.}~\cite{jcp2020}.

We show that progressive dehydration drives ion association, producing two distinct transitions. The first, at an average Zn coordination number $Z=2$, marks a sharp crossover from simple Zn–Cl association to the formation of branched Zn–Cl clusters in which chlorides bridge multiple Zn$^{2+}$ ions. The second is the onset of gelation, the appearance of a percolating Zn–Cl network, predicted near $Z\approx 3$ and confirmed by a critical power-law cluster-size distribution in our MD simulations.

\paragraph*{Molecular Dynamics Simulations.}
We used MD trajectories from our previous work \cite{Cao_PRXEnergy}, performed for aqueous ZnCl$_2$ electrolytes at six molalities, from 1.05 m (Zn:O = 1:53) to 30 m (Zn:O = 1:1.85). The simulations employed a machine-learning interatomic potential (MLIP) trained using DeePMD-kit \cite{deepmd} with density functional theory (DFT) data from Quantum Espresso \cite{QE_Giannozzi_2009,QE_Giannozzi_2017} using the SCAN functional \cite{SCAN_PhysRevLett}. Training data were generated by an active-learning workflow (DP-GEN) \cite{DPGEN}. The resulting MLIP achieved RMS errors of $6.39\!\times\!10^{-4}$ eV/atom for energies and $7.23\!\times\!10^{-4}$ eV/\AA\ for forces.

MD simulations were carried out in LAMMPS \cite{LAMMPS} using a $\sim$25\,$\times$\,25\,$\times$\,25 Å$^3$ box containing 1600–2000 atoms, with periodic boundary conditions. The $NpT$ ensemble at 333 K and 1 bar was used. At 333 K, our SCAN MLIP reproduces the structure of ambient water in the literature~\cite{333K_pnas.1712499114,333K_PhysRevLett.101.017801}. The first 2 ns of each trajectory were discarded as equilibration, and three 100 ns trajectories were collected for each concentration with a 0.5 fs timestep. Comparison of structural factors with X-ray scattering data showed good agreement \cite{Cao_PRXEnergy}, validating the MLIP across all concentrations.

\begin{figure} 
    \centering
    \includegraphics[width=\linewidth]{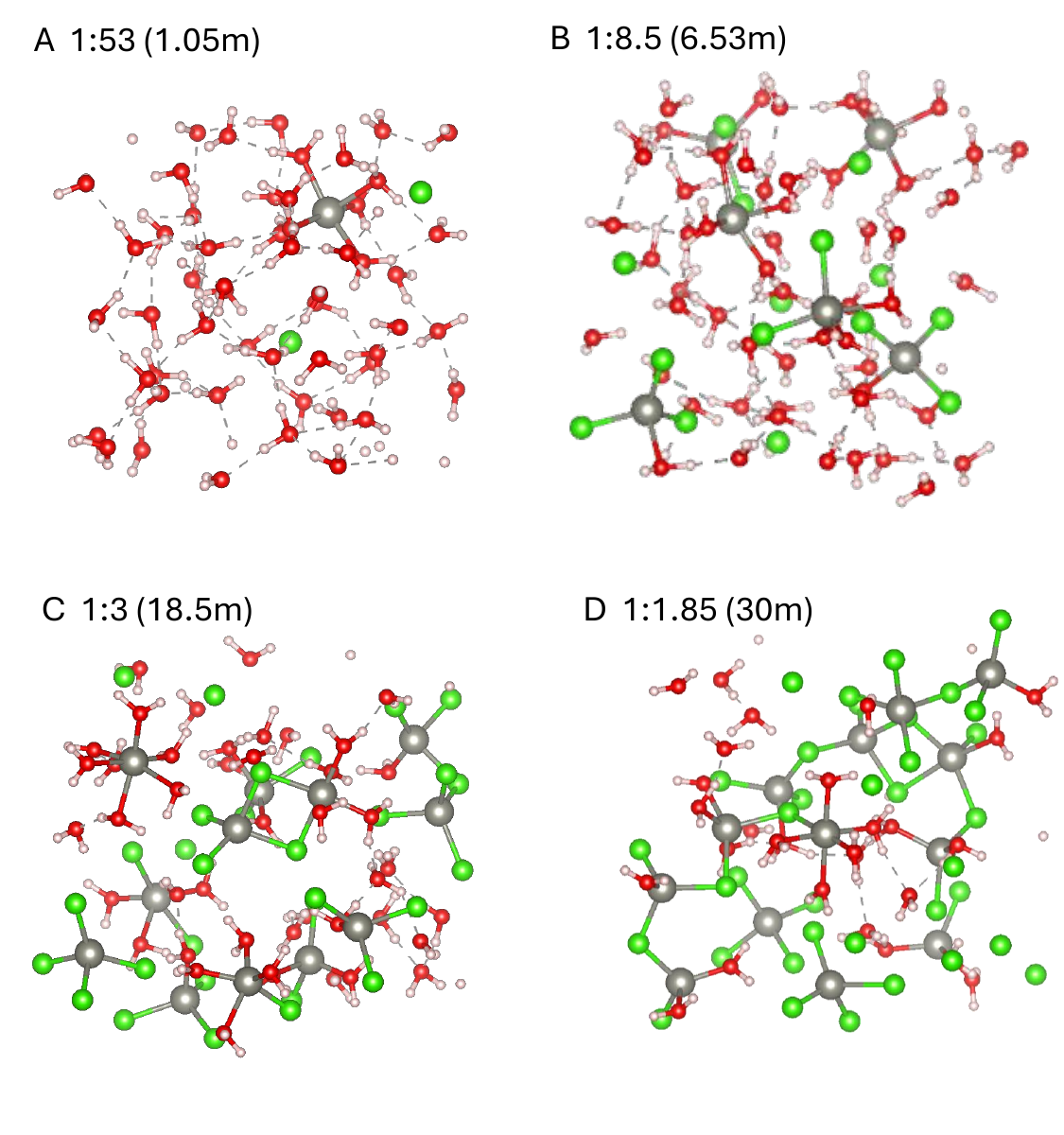}  
    \caption{MD simulation snapshots taken at different salt concentrations. A: salt-to-water molecular ratio $n_s=1:53$ (molality $1.05$ m); B: $n_s=1:8.5$ ($6.53$ m); C: $n_s=1:3$ ($18.5$ m); D: $n_s=1:1.85$ ($30$ m).  }
    \label{fig:snaps}
\end{figure}

Simulation snapshots taken at various concentrations are shown in Figure \ref{fig:snaps}. The trajectories reveal that the first solvation shell (FSS) of Zn$^{2+}$ comprises both H$_2$O and Cl$^-$ ligands, with coordination numbers varying from 4 to 6. Fully hydrated Zn(H$_2$O)$_6^{2+}$ is octahedral, while ZnCl$_4^{2-}$ is tetrahedral. Intermediate complexes Zn(H$_2$O)$_x$Cl$_y$, $x\!+\!y=4$–6, form distorted polyhedra whose chloride content increases with concentration, reflecting progressive ligand substitution. Because each Cl$^-$ can bridge two Zn$^{2+}$ centers, partially chlorinated solvation shells interconnect via shared chlorides, forming extended Zn–Cl aggregates~\cite{JI202199,Choi_ionagg}. In the WIS regime, these aggregates merge into polymer-like ion networks spanning nanometer scales. This evolution from discrete complexes to extended networks controls ion transport and marks the onset of gelation.

\paragraph*{Analytical Model.}

\begin{figure} 
    \centering
    \includegraphics[width=0.8\linewidth]{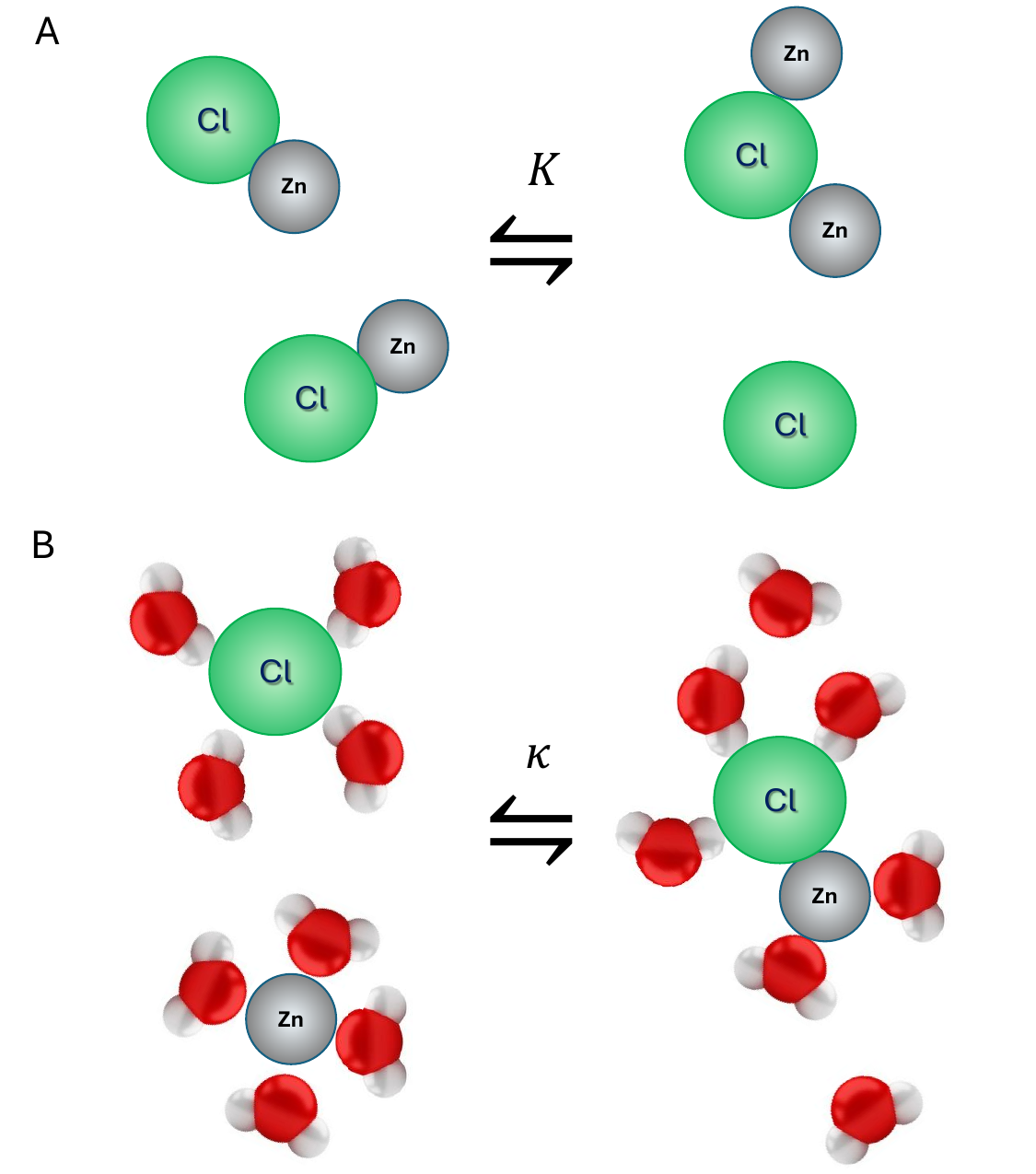}  
    \caption{Schematic representation of the key processes in the model. A: Bridging reaction. B: Hydration-dependent association of Zn$^{2+}$ and Cl$^{-}$ ions. }
    \label{fig:model}
\end{figure}

In our model, each Zn$^{2+}$ provides four equivalent coordination sites for Cl$^-$. Chlorides may be terminal (bound to one Zn) or bridging (shared by two). The central variables are the free-chloride fraction $\alpha$, the mean Zn-Cl coordination number $Z$, the bridging fraction $p$, and the salt-to-water molecular ratio $n_s=C_s/C_w$, where  $C_s$ and $C_w$ represent water and salt concentrations, respectively. Terminal Cl$^-$ contributes one Zn–Cl bond, while a bridge contributes two, giving the coordination balance
\begin{equation}
Z = 2(1-\alpha)(1+p).
\label{eq:Z}
\end{equation}
The conversion of two terminals into a bridge releases one free chloride, as shown in Fig. \ref{fig:model}A,  yielding the equilibrium condition:
\begin{equation}
\frac{p}{K(1-p)^2} = \frac{1-\alpha}{\alpha},
\label{eq:bridge}
\end{equation}
where the dimensionless constant $K\!\ll\!1$ encodes the electrostatic and entropic penalties of bridging.
\begin{figure}[b]
    \centering
    \includegraphics[width=\linewidth]{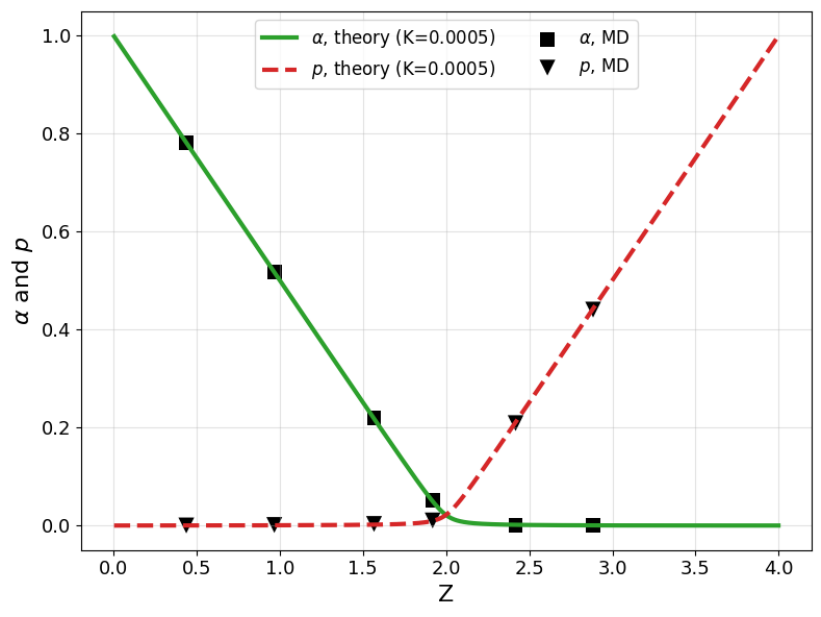}
    \caption{Theoretical relationships between $p(Z)$ (dashed) and $\alpha(Z)$ (solid) compared with MD data (symbols) for $K=0.0005$.}
    \label{fig:Zplots}
\end{figure}

Combining Eqs.~(\ref{eq:Z})–(\ref{eq:bridge}) gives
\begin{align}
\alpha(p)&=\frac{1}{1+p/[K(1-p)^2]},\\
Z(p)&=\frac{2(1+p)}{1+K(1-p)^2/p}.
\label{eq:Zp}
\end{align}
As shown in Fig.~\ref{fig:Zplots}, these predictions quantitatively match MD data for $K=5\times10^{-4}$.

Two regimes emerge with a sharp crossover at $Z=2$.  
For $Z<2$, $\alpha\!\approx\!1-Z/2$ and $p\!\approx\!0$, corresponding to dissociated ions and small clusters.  
For $Z>2$, $\alpha\!\approx\!0$ and $p\!\approx\!Z/2-1$, signaling extensive aggregation.  
This reflects the large free-energy cost of Zn–Cl–Zn bridging: at low $Z$, the system favors isolated ions and single-Zn clusters; but once $Z=2$, chloride scarcity forces the system to use bridges, driving rapid aggregation. We therefore identify $Z=2$ as the \textit{aggregation threshold}.

The preceding argument explains why a sharp crossover occurs at $Z=2$, but it does not by itself guarantee that the solution can reach this point prior to becoming saturated. To illustrate the limitation, first consider a naive model in which binding between free chloride and vacant Zn sites is described by a single association constant $K'$:
\begin{align}
    K'\,[\mathrm{Cl^-}][\mathrm{Zn\!-\!site}] &= [\mathrm{Cl}_\mathrm{t}],\\
    K'\,C_s(4-Z) &= \frac{1-\alpha}{\alpha}.
\end{align}
Here, the dissociation fraction $\alpha$ decreases slowly,  as $1/C_s$. This implies that the exponentially small value  $\alpha\sim 1/K$ expected near $Z=2$, would be achieved at  $C_s\sim K/K'$. In other words, the salt concentration should be increased by an additional factor $K$ compared to the onset of association, which yields an unrealistically high threshold,  $C_s> 1000 m$.   

\begin{figure} 
    \centering
    \includegraphics[width=\linewidth]{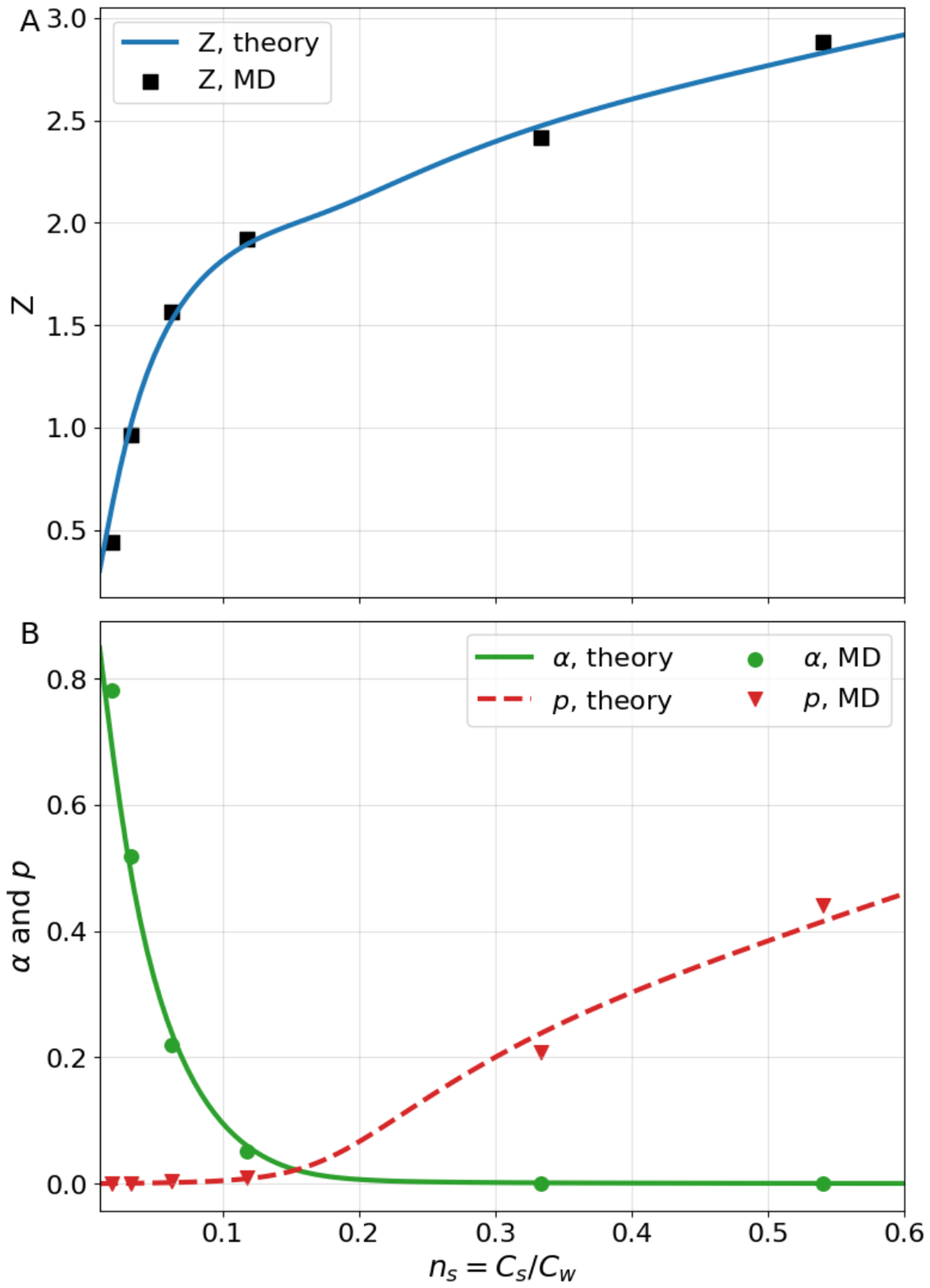}  
    \caption{Comparison of theoretical dependencies of the mean Zn-Cl coordination number $Z$ (panel A), the mean bridging probability $p$, and the Cl$^-$ dissociation fraction $\alpha$ with MD simulations (panel B).}
    \label{fig:mainplot}
\end{figure}

The naive association model does not take into account that at high enough salt concentration the availability of water molecules for the hydration of free ions becomes a limiting factor. To capture the separate scarcity of hydrogen donors (for anion hydration) and oxygen acceptors (for cation hydration), we introduce two dimensionless coefficients, $\gamma_{\mathrm{H}}$ and $\gamma_{\mathrm{O}}$, the solvent-site activity coefficients for hydrogen and oxygen sites, respectively. They play the role of activity coefficients for the respective solvent sites: $\gamma=1$ indicates ideal availability; $\gamma<1$ indicates scarcity. 

We begin with O-sites. A fully hydrated Zn$^{2+}$ ion has $s_0=6$ oxygen ligands in the first shell, while a tetrahedral ZnCl$_4$ species has none. Thus, each Zn--Cl bond, on average, blocks $s\approx 3/2$ oxygen-binding sites. If $(1-\gamma_{\mathrm{O}})C_w$ denotes the population of O-sites that are effectively sequestered, a mass-action balance leads to
\begin{equation}
   K_O\,\gamma_{\mathrm{O}}\, C_w \Big( (s_0-sZ)C_s-(1-\gamma_{\mathrm{O}})C_w \Big) =(1-\gamma_{\mathrm{O}})C_w.
\end{equation}
Here association constant $K_O$ characterizes the affinity of an O-site of the water molecule for a Zn$^{2+}$ ion. A similar relation can be written for $\gamma_{\mathrm{H}}$, assuming that hydration of a free Cl$^-$ requires $h_0$ H-donor sites and that each Zn--Cl bond frees $h$ of them. Because the Cl$^-$ hydration shell is less structured than that of Zn$^{2+}$, we take $h_0=8$ and treat $h$ as a fitting parameter. From the above balance equation, one obtains
\begin{align}
   \gamma_{\mathrm{O}}-\frac{\epsilon_O}{\gamma_{\mathrm{O}}}&\approx 1 + n_s(s Z - s_0),
\label{eq:O}\\
\gamma_{\mathrm{H}}-\frac{\epsilon_H}{\gamma_{\mathrm{H}}}&\approx 1 + n_s\!\left(\frac{hZ}{2}- h_0\right),
\label{eq:H} 
\end{align}
with $\epsilon_O=1/(K_O C_s)\ll1$ and $\epsilon_H$ the analogous small parameter for H-sites.

Associating a Cl$^-$ with a vacant Zn site releases $s$ oxygen sites and $h$ hydrogen sites, as schematically shown in Fig. \ref{fig:model}B:
\begin{equation}
\mathrm{Zn\!-\!site} + \mathrm{Cl^-} \rightleftharpoons \mathrm{Zn\!-\!Cl}+ s\,\mathrm{O}+ h\,\mathrm{H}.
\end{equation}
Mass action equilibrium gives
\begin{equation}
\frac{1-\alpha}{\alpha}
= \frac{\kappa\, n_s}{\gamma_{\mathrm{O}}^{s}\gamma_{\mathrm{H}}^{h}}\,\frac{4-Z}{1-p},
\label{eq:assoc}
\end{equation}
with $\kappa$ is the reduced association constant for this reaction, which is made dimensionless by adsorbing $C_s$ within its definition.  Combining Eqs.~(\ref{eq:assoc}), (\ref{eq:bridge}), and~(\ref{eq:Zp}) yields the full dependence of $Z$, $p$, and $\alpha$ on $n_s$. Figure~\ref{fig:mainplot} compares the theoretical predictions with MD results using $K=0.0005$, $\kappa=3.0$, and $h=5.2$.

\paragraph*{Onset of Gelation.}

Our system belongs to the broad class of gelation and percolation models for branched-monomer networks. Classical FS theory predicts gelation for monomers of functionality $f$ when the bond-formation probability exceeds $p_b > p_c = 1/(f-1)$. For ZnCl$_2$, the relevant functionality is $f=4$, and the effective Zn--Zn bonding probability is $Zp/4$, which yields the gelation condition
\begin{align}
&Z(Z-2) > 8p_c, \\
&Z_c = 1 + \sqrt{\,1 + 8p_c\,}.
\end{align}
Using the FS value $p_c = 1/3$ gives $Z_c \approx 2.91$~\cite{Flory1941,Stockmayer1944}. However, FS theory is a mean-field approximation to the full percolation problem, as it neglects loop formation and thus cannot capture the correct critical behavior~\cite{perc2018introduction}. A more realistic estimate is obtained from bond percolation on the diamond lattice (also with $f=4$), which has $p_c \approx 0.39$, implying $Z_c \approx 3.03$~\cite{perc3D}. At the highest concentration studied ($n_s = 1{:}1.85$, or 30 m), the mean coordination number $Z$ approaches this critical value. Below we verify the proximity to criticality by examining the cluster-size distribution.

\begin{figure} 
    \centering
    \includegraphics[width=\linewidth]{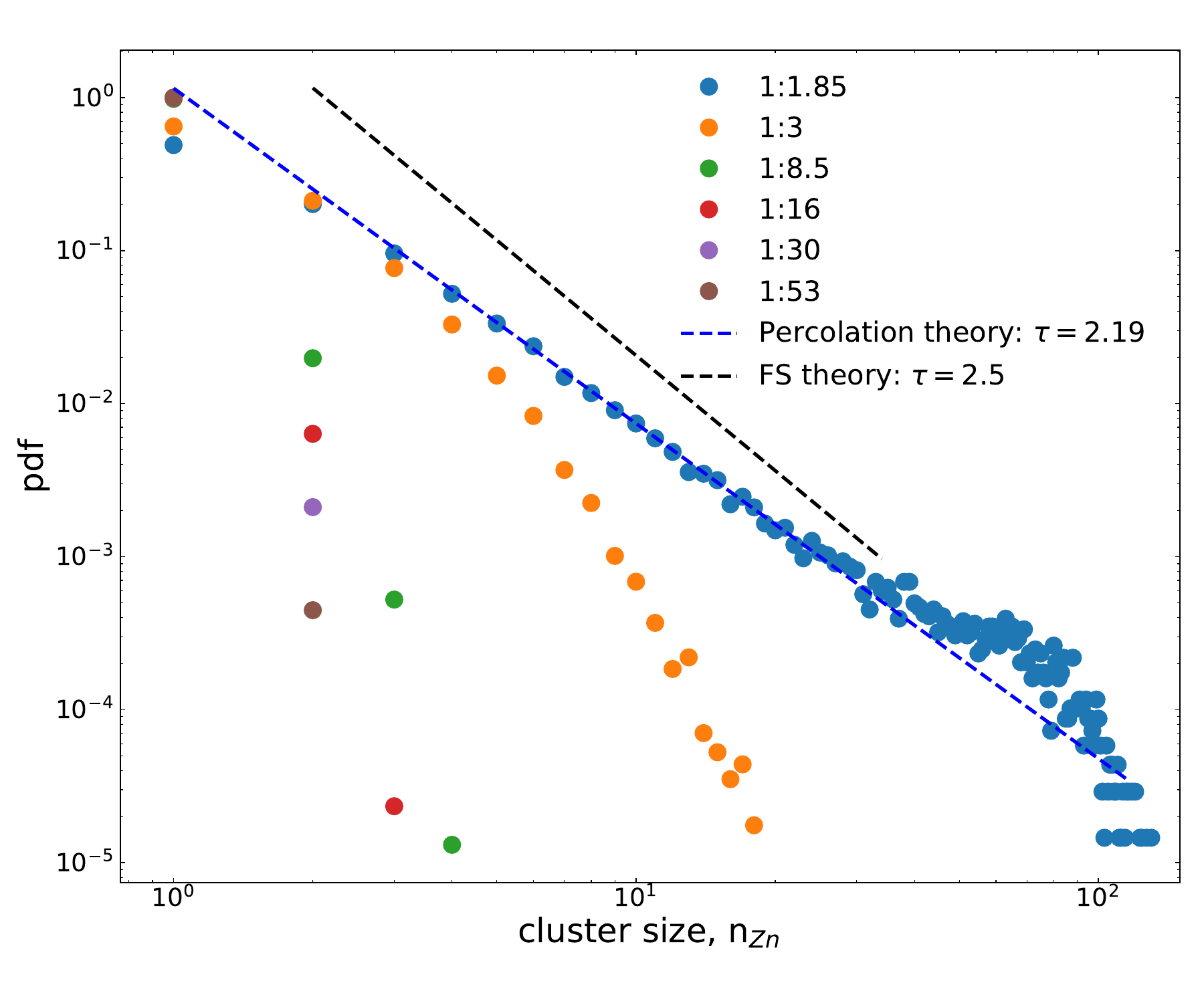}  
    \caption{Cluster population distributions from MD simulations for Zn-Cl aggregates at different concentrations, plotted on log-log scale. The markers show probability density function (pdf) vs. the number of Zn$^{2+}$ ions per cluster. In the 30 m electrolyte (Zn:O = 1:1.85), the distribution follow the  power-law behavior $f\sim n_{Zn}^{-\tau}$, with expected critical exponent  of $ \tau=2.19$ for percolation theory in 3D. This behavior is  distinctly different from loop-free result of FS theory, $\tau=5/2$  }
    \label{fig:loglog-plot}
\end{figure}

The gelation point corresponds to the emergence of an infinite cluster; near this transition, the cluster-size distribution obeys the universal scaling law $f(n) \sim n^{-\tau}$, where $\tau$ is the Fisher exponent \cite{perc2018introduction}. The loopless FS theory predicts $\tau_{\mathrm{FS}} = 5/2$ \cite{Flory1941,Stockmayer1944}, whereas the 3D percolation universality class has a lower exponent, $\tau \approx 2.19$ \cite{perc2018introduction,perc3D}. Figure~\ref{fig:loglog-plot} shows the cluster-size distributions obtained from MD simulations at various concentrations. At the highest concentration, the distribution becomes extremely broad, ranging from isolated Zn$^{2+}$ ions to aggregates containing $\sim$130 Zn$^{2+}$. The log--log representation exhibits a clear power-law in good agreement with the critical exponent expected for 3D percolation $\tau = 2.19$, and clearly distinct from the exponent $\tau=5/2$ predicted by the mean-field  FS theory. This scaling behavior confirms the system's proximity to the percolation threshold and, indirectly, the crucial role of loop formation in determining the cluster statistics.

In summary, our combined molecular dynamics simulations and analytic modeling demonstrate that progressive dehydration is the key microscopic driver of ion aggregation and gelation in concentrated ZnCl$_2$ solutions.
As water becomes scarce, chloride ligands increasingly bridge neighboring Zn$^{2+}$ centers, leading to a hierarchy of structural transitions. The first occurs at an average coordination number $Z=2$, marking the crossover from isolated ionic complexes ZnCl$_x$ to branched clusters containing multiple Zn$^{2+}$ ions.
The second transition, at $Z\approx3$, corresponds to the emergence of a system-spanning Zn–Cl network, a percolation threshold analogous to the sol–gel transition in classical polymer systems.
The resulting power-law cluster-size distribution observed in simulations provides clear evidence of near-critical behavior consistent with three-dimensional percolation theory.

These findings establish a direct, quantitative connection between local coordination chemistry and emergent mesoscale connectivity in aqueous electrolytes.
They also highlight dehydration-induced ion bridging as a universal mechanism governing the structural evolution of water-in-salt systems. The analytical model provides insights into the charge carrier characteristics, thermodynamic properties and ion transport properties (such as conductivity and Zn$^{2+}$ transference number) of the ZnCl$_2$ eletrolyte at different concentrations~\cite{Cao_PRXEnergy}.
Beyond ZnCl$_2$, this framework offers a general physical basis for understanding the interplay between solvation, ion association, and gelation in highly concentrated electrolytes and other complex ionic liquids of technological importance.

\paragraph{Acknowledgments} This research has been conducted and used resources of the Theory and Computational Facility of the Center for Functional Nanomaterials, which is U.S. Department of Energy (DOE) Office of Science User Facilities operated for the DOE Office of Science by Brookhaven National Laboratory under Contract No. DE-SC0012704. This  was partially funded by the U.S. Department of Energy, Office of Science, Basic Energy Sciences, through Contract No. DE-SC0012704, under the Chemical and Materials Sciences to Advance Clean-Energy Technologies and Transform Manufacturing (CEM) program.


\begin{thebibliography}{20}%
\makeatletter
\providecommand \@ifxundefined [1]{%
 \@ifx{#1\undefined}
}%
\providecommand \@ifnum [1]{%
 \ifnum #1\expandafter \@firstoftwo
 \else \expandafter \@secondoftwo
 \fi
}%
\providecommand \@ifx [1]{%
 \ifx #1\expandafter \@firstoftwo
 \else \expandafter \@secondoftwo
 \fi
}%
\providecommand \natexlab [1]{#1}%
\providecommand \enquote  [1]{``#1''}%
\providecommand \bibnamefont  [1]{#1}%
\providecommand \bibfnamefont [1]{#1}%
\providecommand \citenamefont [1]{#1}%
\providecommand \href@noop [0]{\@secondoftwo}%
\providecommand \href [0]{\begingroup \@sanitize@url \@href}%
\providecommand \@href[1]{\@@startlink{#1}\@@href}%
\providecommand \@@href[1]{\endgroup#1\@@endlink}%
\providecommand \@sanitize@url [0]{\catcode `\\12\catcode `\$12\catcode `\&12\catcode `\#12\catcode `\^12\catcode `\_12\catcode `\%12\relax}%
\providecommand \@@startlink[1]{}%
\providecommand \@@endlink[0]{}%
\providecommand \url  [0]{\begingroup\@sanitize@url \@url }%
\providecommand \@url [1]{\endgroup\@href {#1}{\urlprefix }}%
\providecommand \urlprefix  [0]{URL }%
\providecommand \Eprint [0]{\href }%
\providecommand \doibase [0]{https://doi.org/}%
\providecommand \selectlanguage [0]{\@gobble}%
\providecommand \bibinfo  [0]{\@secondoftwo}%
\providecommand \bibfield  [0]{\@secondoftwo}%
\providecommand \translation [1]{[#1]}%
\providecommand \BibitemOpen [0]{}%
\providecommand \bibitemStop [0]{}%
\providecommand \bibitemNoStop [0]{.\EOS\space}%
\providecommand \EOS [0]{\spacefactor3000\relax}%
\providecommand \BibitemShut  [1]{\csname bibitem#1\endcsname}%
\let\auto@bib@innerbib\@empty
\bibitem [{\citenamefont {Ji}(2021)}]{JI202199}%
  \BibitemOpen
  \bibfield  {author} {\bibinfo {author} {\bibfnamefont {X.}~\bibnamefont {Ji}},\ }\bibfield  {title} {\bibinfo {title} {A perspective of zncl2 electrolytes: The physical and electrochemical properties},\ }\href {https://doi.org/https://doi.org/10.1016/j.esci.2021.10.004} {\bibfield  {journal} {\bibinfo  {journal} {eScience}\ }\textbf {\bibinfo {volume} {1}},\ \bibinfo {pages} {99} (\bibinfo {year} {2021})}\BibitemShut {NoStop}%
\bibitem [{\citenamefont {Wang}\ \emph {et~al.}(2018{\natexlab{a}})\citenamefont {Wang}, \citenamefont {Borodin}, \citenamefont {Gao}, \citenamefont {Fan}, \citenamefont {Sun}, \citenamefont {Han}, \citenamefont {Faraone}, \citenamefont {Dura}, \citenamefont {Xu},\ and\ \citenamefont {Wang}}]{wang2018highly}%
  \BibitemOpen
  \bibfield  {author} {\bibinfo {author} {\bibfnamefont {F.}~\bibnamefont {Wang}}, \bibinfo {author} {\bibfnamefont {O.}~\bibnamefont {Borodin}}, \bibinfo {author} {\bibfnamefont {T.}~\bibnamefont {Gao}}, \bibinfo {author} {\bibfnamefont {X.}~\bibnamefont {Fan}}, \bibinfo {author} {\bibfnamefont {W.}~\bibnamefont {Sun}}, \bibinfo {author} {\bibfnamefont {F.}~\bibnamefont {Han}}, \bibinfo {author} {\bibfnamefont {A.}~\bibnamefont {Faraone}}, \bibinfo {author} {\bibfnamefont {J.~A.}\ \bibnamefont {Dura}}, \bibinfo {author} {\bibfnamefont {K.}~\bibnamefont {Xu}},\ and\ \bibinfo {author} {\bibfnamefont {C.}~\bibnamefont {Wang}},\ }\bibfield  {title} {\bibinfo {title} {Highly reversible zinc metal anode for aqueous batteries},\ }\href@noop {} {\bibfield  {journal} {\bibinfo  {journal} {Nature materials}\ }\textbf {\bibinfo {volume} {17}},\ \bibinfo {pages} {543} (\bibinfo {year} {2018}{\natexlab{a}})}\BibitemShut {NoStop}%
\bibitem [{\citenamefont {Cao}\ \emph {et~al.}(2025)\citenamefont {Cao}, \citenamefont {Kingan}, \citenamefont {Hill}, \citenamefont {Kuang}, \citenamefont {Wang}, \citenamefont {Zhang}, \citenamefont {Carbone}, \citenamefont {van Dam}, \citenamefont {Yoo}, \citenamefont {Yan}, \citenamefont {Takeuchi}, \citenamefont {Takeuchi}, \citenamefont {Wu}, \citenamefont {Abeykoon}, \citenamefont {Marschilok},\ and\ \citenamefont {Lu}}]{Cao_PRXEnergy}%
  \BibitemOpen
  \bibfield  {author} {\bibinfo {author} {\bibfnamefont {C.}~\bibnamefont {Cao}}, \bibinfo {author} {\bibfnamefont {A.}~\bibnamefont {Kingan}}, \bibinfo {author} {\bibfnamefont {R.~C.}\ \bibnamefont {Hill}}, \bibinfo {author} {\bibfnamefont {J.}~\bibnamefont {Kuang}}, \bibinfo {author} {\bibfnamefont {L.}~\bibnamefont {Wang}}, \bibinfo {author} {\bibfnamefont {C.}~\bibnamefont {Zhang}}, \bibinfo {author} {\bibfnamefont {M.~R.}\ \bibnamefont {Carbone}}, \bibinfo {author} {\bibfnamefont {H.}~\bibnamefont {van Dam}}, \bibinfo {author} {\bibfnamefont {S.}~\bibnamefont {Yoo}}, \bibinfo {author} {\bibfnamefont {S.}~\bibnamefont {Yan}}, \bibinfo {author} {\bibfnamefont {E.~S.}\ \bibnamefont {Takeuchi}}, \bibinfo {author} {\bibfnamefont {K.~J.}\ \bibnamefont {Takeuchi}}, \bibinfo {author} {\bibfnamefont {X.}~\bibnamefont {Wu}}, \bibinfo {author} {\bibfnamefont {A.~M.}\ \bibnamefont {Abeykoon}}, \bibinfo {author} {\bibfnamefont {A.~C.}\ \bibnamefont {Marschilok}},\ and\ \bibinfo {author} {\bibfnamefont
  {D.}~\bibnamefont {Lu}},\ }\bibfield  {title} {\bibinfo {title} {Resolving the solvation structure and transport properties of aqueous zinc electrolytes from salt-in-water to water-in-salt using neural network potential},\ }\href {https://doi.org/10.1103/PRXEnergy.4.023004} {\bibfield  {journal} {\bibinfo  {journal} {PRX Energy}\ }\textbf {\bibinfo {volume} {4}},\ \bibinfo {pages} {023004} (\bibinfo {year} {2025})}\BibitemShut {NoStop}%
\bibitem [{\citenamefont {McEldrew}\ \emph {et~al.}(2020)\citenamefont {McEldrew}, \citenamefont {Goodwin}, \citenamefont {Bi}, \citenamefont {Bazant},\ and\ \citenamefont {Kornyshev}}]{jcp2020}%
  \BibitemOpen
  \bibfield  {author} {\bibinfo {author} {\bibfnamefont {M.}~\bibnamefont {McEldrew}}, \bibinfo {author} {\bibfnamefont {Z.~A.~H.}\ \bibnamefont {Goodwin}}, \bibinfo {author} {\bibfnamefont {S.}~\bibnamefont {Bi}}, \bibinfo {author} {\bibfnamefont {M.~Z.}\ \bibnamefont {Bazant}},\ and\ \bibinfo {author} {\bibfnamefont {A.~A.}\ \bibnamefont {Kornyshev}},\ }\bibfield  {title} {\bibinfo {title} {Theory of ion aggregation and gelation in super-concentrated electrolytes},\ }\href {https://doi.org/10.1063/5.0006197} {\bibfield  {journal} {\bibinfo  {journal} {The Journal of Chemical Physics}\ }\textbf {\bibinfo {volume} {152}},\ \bibinfo {pages} {234506} (\bibinfo {year} {2020})}\BibitemShut {NoStop}%
\bibitem [{\citenamefont {McEldrew}\ \emph {et~al.}(2021)\citenamefont {McEldrew}, \citenamefont {Goodwin}, \citenamefont {Bi}, \citenamefont {Kornyshev},\ and\ \citenamefont {Bazant}}]{McEldrew_2021}%
  \BibitemOpen
  \bibfield  {author} {\bibinfo {author} {\bibfnamefont {M.}~\bibnamefont {McEldrew}}, \bibinfo {author} {\bibfnamefont {Z.~A.~H.}\ \bibnamefont {Goodwin}}, \bibinfo {author} {\bibfnamefont {S.}~\bibnamefont {Bi}}, \bibinfo {author} {\bibfnamefont {A.~A.}\ \bibnamefont {Kornyshev}},\ and\ \bibinfo {author} {\bibfnamefont {M.~Z.}\ \bibnamefont {Bazant}},\ }\bibfield  {title} {\bibinfo {title} {Ion clusters and networks in water-in-salt electrolytes},\ }\href {https://doi.org/10.1149/1945-7111/abf975} {\bibfield  {journal} {\bibinfo  {journal} {Journal of The Electrochemical Society}\ }\textbf {\bibinfo {volume} {168}},\ \bibinfo {pages} {050514} (\bibinfo {year} {2021})}\BibitemShut {NoStop}%
\bibitem [{\citenamefont {Goodwin}\ \emph {et~al.}(2023)\citenamefont {Goodwin}, \citenamefont {McEldrew}, \citenamefont {Kozinsky},\ and\ \citenamefont {Bazant}}]{PRXEnergy2023}%
  \BibitemOpen
  \bibfield  {author} {\bibinfo {author} {\bibfnamefont {Z.~A.}\ \bibnamefont {Goodwin}}, \bibinfo {author} {\bibfnamefont {M.}~\bibnamefont {McEldrew}}, \bibinfo {author} {\bibfnamefont {B.}~\bibnamefont {Kozinsky}},\ and\ \bibinfo {author} {\bibfnamefont {M.~Z.}\ \bibnamefont {Bazant}},\ }\bibfield  {title} {\bibinfo {title} {Theory of cation solvation and ionic association in nonaqueous solvent mixtures},\ }\href {https://doi.org/10.1103/PRXEnergy.2.013007} {\bibfield  {journal} {\bibinfo  {journal} {PRX Energy}\ }\textbf {\bibinfo {volume} {2}},\ \bibinfo {pages} {013007} (\bibinfo {year} {2023})}\BibitemShut {NoStop}%
\bibitem [{\citenamefont {Flory}(1941)}]{Flory1941}%
  \BibitemOpen
  \bibfield  {author} {\bibinfo {author} {\bibfnamefont {P.~J.}\ \bibnamefont {Flory}},\ }\bibfield  {title} {\bibinfo {title} {Molecular size distribution in three dimensional polymers. i. gelation1},\ }\href {https://doi.org/10.1021/ja01856a061} {\bibfield  {journal} {\bibinfo  {journal} {Journal of the American Chemical Society}\ }\textbf {\bibinfo {volume} {63}},\ \bibinfo {pages} {3083} (\bibinfo {year} {1941})}\BibitemShut {NoStop}%
\bibitem [{\citenamefont {Stockmayer}(1944)}]{Stockmayer1944}%
  \BibitemOpen
  \bibfield  {author} {\bibinfo {author} {\bibfnamefont {W.~H.}\ \bibnamefont {Stockmayer}},\ }\bibfield  {title} {\bibinfo {title} {Theory of molecular size distribution and gel formation in branched polymers ii. general cross linking},\ }\href {https://doi.org/10.1063/1.1723922} {\bibfield  {journal} {\bibinfo  {journal} {The Journal of Chemical Physics}\ }\textbf {\bibinfo {volume} {12}},\ \bibinfo {pages} {125} (\bibinfo {year} {1944})}\BibitemShut {NoStop}%
\bibitem [{\citenamefont {Tanaka}\ and\ \citenamefont {Stockmayer}(1994)}]{Tanaka1994}%
  \BibitemOpen
  \bibfield  {author} {\bibinfo {author} {\bibfnamefont {F.}~\bibnamefont {Tanaka}}\ and\ \bibinfo {author} {\bibfnamefont {W.~H.}\ \bibnamefont {Stockmayer}},\ }\bibfield  {title} {\bibinfo {title} {Thermoreversible gelation with junctions of variable multiplicity},\ }\href {https://doi.org/https://doi.org/10.1002/masy.19940810119} {\bibfield  {journal} {\bibinfo  {journal} {Macromolecular Symposia}\ }\textbf {\bibinfo {volume} {81}},\ \bibinfo {pages} {171} (\bibinfo {year} {1994})}\BibitemShut {NoStop}%
\bibitem [{\citenamefont {Wang}\ \emph {et~al.}(2018{\natexlab{b}})\citenamefont {Wang}, \citenamefont {Zhang}, \citenamefont {Han},\ and\ \citenamefont {E}}]{deepmd}%
  \BibitemOpen
  \bibfield  {author} {\bibinfo {author} {\bibfnamefont {H.}~\bibnamefont {Wang}}, \bibinfo {author} {\bibfnamefont {L.}~\bibnamefont {Zhang}}, \bibinfo {author} {\bibfnamefont {J.}~\bibnamefont {Han}},\ and\ \bibinfo {author} {\bibfnamefont {W.}~\bibnamefont {E}},\ }\bibfield  {title} {\bibinfo {title} {Deepmd-kit: A deep learning package for many-body potential energy representation and molecular dynamics},\ }\href {https://doi.org/https://doi.org/10.1016/j.cpc.2018.03.016} {\bibfield  {journal} {\bibinfo  {journal} {Computer Physics Communications}\ }\textbf {\bibinfo {volume} {228}},\ \bibinfo {pages} {178} (\bibinfo {year} {2018}{\natexlab{b}})}\BibitemShut {NoStop}%
\bibitem [{\citenamefont {Giannozzi}\ \emph {et~al.}(2009)\citenamefont {Giannozzi}, \citenamefont {Baroni}, \citenamefont {Bonini}, \citenamefont {Calandra}, \citenamefont {Car}, \citenamefont {Cavazzoni}, \citenamefont {Ceresoli}, \citenamefont {Chiarotti}, \citenamefont {Cococcioni}, \citenamefont {Dabo}, \citenamefont {Dal~Corso}, \citenamefont {de~Gironcoli}, \citenamefont {Fabris}, \citenamefont {Fratesi}, \citenamefont {Gebauer}, \citenamefont {Gerstmann}, \citenamefont {Gougoussis}, \citenamefont {Kokalj}, \citenamefont {Lazzeri}, \citenamefont {Martin-Samos}, \citenamefont {Marzari}, \citenamefont {Mauri}, \citenamefont {Mazzarello}, \citenamefont {Paolini}, \citenamefont {Pasquarello}, \citenamefont {Paulatto}, \citenamefont {Sbraccia}, \citenamefont {Scandolo}, \citenamefont {Sclauzero}, \citenamefont {Seitsonen}, \citenamefont {Smogunov}, \citenamefont {Umari},\ and\ \citenamefont {Wentzcovitch}}]{QE_Giannozzi_2009}%
  \BibitemOpen
  \bibfield  {author} {\bibinfo {author} {\bibfnamefont {P.}~\bibnamefont {Giannozzi}}, \bibinfo {author} {\bibfnamefont {S.}~\bibnamefont {Baroni}}, \bibinfo {author} {\bibfnamefont {N.}~\bibnamefont {Bonini}}, \bibinfo {author} {\bibfnamefont {M.}~\bibnamefont {Calandra}}, \bibinfo {author} {\bibfnamefont {R.}~\bibnamefont {Car}}, \bibinfo {author} {\bibfnamefont {C.}~\bibnamefont {Cavazzoni}}, \bibinfo {author} {\bibfnamefont {D.}~\bibnamefont {Ceresoli}}, \bibinfo {author} {\bibfnamefont {G.~L.}\ \bibnamefont {Chiarotti}}, \bibinfo {author} {\bibfnamefont {M.}~\bibnamefont {Cococcioni}}, \bibinfo {author} {\bibfnamefont {I.}~\bibnamefont {Dabo}}, \bibinfo {author} {\bibfnamefont {A.}~\bibnamefont {Dal~Corso}}, \bibinfo {author} {\bibfnamefont {S.}~\bibnamefont {de~Gironcoli}}, \bibinfo {author} {\bibfnamefont {S.}~\bibnamefont {Fabris}}, \bibinfo {author} {\bibfnamefont {G.}~\bibnamefont {Fratesi}}, \bibinfo {author} {\bibfnamefont {R.}~\bibnamefont {Gebauer}}, \bibinfo {author} {\bibfnamefont
  {U.}~\bibnamefont {Gerstmann}}, \bibinfo {author} {\bibfnamefont {C.}~\bibnamefont {Gougoussis}}, \bibinfo {author} {\bibfnamefont {A.}~\bibnamefont {Kokalj}}, \bibinfo {author} {\bibfnamefont {M.}~\bibnamefont {Lazzeri}}, \bibinfo {author} {\bibfnamefont {L.}~\bibnamefont {Martin-Samos}}, \bibinfo {author} {\bibfnamefont {N.}~\bibnamefont {Marzari}}, \bibinfo {author} {\bibfnamefont {F.}~\bibnamefont {Mauri}}, \bibinfo {author} {\bibfnamefont {R.}~\bibnamefont {Mazzarello}}, \bibinfo {author} {\bibfnamefont {S.}~\bibnamefont {Paolini}}, \bibinfo {author} {\bibfnamefont {A.}~\bibnamefont {Pasquarello}}, \bibinfo {author} {\bibfnamefont {L.}~\bibnamefont {Paulatto}}, \bibinfo {author} {\bibfnamefont {C.}~\bibnamefont {Sbraccia}}, \bibinfo {author} {\bibfnamefont {S.}~\bibnamefont {Scandolo}}, \bibinfo {author} {\bibfnamefont {G.}~\bibnamefont {Sclauzero}}, \bibinfo {author} {\bibfnamefont {A.~P.}\ \bibnamefont {Seitsonen}}, \bibinfo {author} {\bibfnamefont {A.}~\bibnamefont {Smogunov}}, \bibinfo {author}
  {\bibfnamefont {P.}~\bibnamefont {Umari}},\ and\ \bibinfo {author} {\bibfnamefont {R.~M.}\ \bibnamefont {Wentzcovitch}},\ }\bibfield  {title} {\bibinfo {title} {Quantum espresso: a modular and open-source software project for quantum simulations of materials},\ }\href {https://doi.org/10.1088/0953-8984/21/39/395502} {\bibfield  {journal} {\bibinfo  {journal} {Journal of Physics: Condensed Matter}\ }\textbf {\bibinfo {volume} {21}},\ \bibinfo {pages} {395502} (\bibinfo {year} {2009})}\BibitemShut {NoStop}%
\bibitem [{\citenamefont {Giannozzi}\ \emph {et~al.}(2017)\citenamefont {Giannozzi}, \citenamefont {Andreussi}, \citenamefont {Brumme}, \citenamefont {Bunau}, \citenamefont {Buongiorno~Nardelli}, \citenamefont {Calandra}, \citenamefont {Car}, \citenamefont {Cavazzoni}, \citenamefont {Ceresoli}, \citenamefont {Cococcioni}, \citenamefont {Colonna}, \citenamefont {Carnimeo}, \citenamefont {Dal~Corso}, \citenamefont {de~Gironcoli}, \citenamefont {Delugas}, \citenamefont {DiStasio}, \citenamefont {Ferretti}, \citenamefont {Floris}, \citenamefont {Fratesi}, \citenamefont {Fugallo}, \citenamefont {Gebauer}, \citenamefont {Gerstmann}, \citenamefont {Giustino}, \citenamefont {Gorni}, \citenamefont {Jia}, \citenamefont {Kawamura}, \citenamefont {Ko}, \citenamefont {Kokalj}, \citenamefont {Küçükbenli}, \citenamefont {Lazzeri}, \citenamefont {Marsili}, \citenamefont {Marzari}, \citenamefont {Mauri}, \citenamefont {Nguyen}, \citenamefont {Nguyen}, \citenamefont {Otero-de-la Roza}, \citenamefont {Paulatto},
  \citenamefont {Poncé}, \citenamefont {Rocca}, \citenamefont {Sabatini}, \citenamefont {Santra}, \citenamefont {Schlipf}, \citenamefont {Seitsonen}, \citenamefont {Smogunov}, \citenamefont {Timrov}, \citenamefont {Thonhauser}, \citenamefont {Umari}, \citenamefont {Vast}, \citenamefont {Wu},\ and\ \citenamefont {Baroni}}]{QE_Giannozzi_2017}%
  \BibitemOpen
  \bibfield  {author} {\bibinfo {author} {\bibfnamefont {P.}~\bibnamefont {Giannozzi}}, \bibinfo {author} {\bibfnamefont {O.}~\bibnamefont {Andreussi}}, \bibinfo {author} {\bibfnamefont {T.}~\bibnamefont {Brumme}}, \bibinfo {author} {\bibfnamefont {O.}~\bibnamefont {Bunau}}, \bibinfo {author} {\bibfnamefont {M.}~\bibnamefont {Buongiorno~Nardelli}}, \bibinfo {author} {\bibfnamefont {M.}~\bibnamefont {Calandra}}, \bibinfo {author} {\bibfnamefont {R.}~\bibnamefont {Car}}, \bibinfo {author} {\bibfnamefont {C.}~\bibnamefont {Cavazzoni}}, \bibinfo {author} {\bibfnamefont {D.}~\bibnamefont {Ceresoli}}, \bibinfo {author} {\bibfnamefont {M.}~\bibnamefont {Cococcioni}}, \bibinfo {author} {\bibfnamefont {N.}~\bibnamefont {Colonna}}, \bibinfo {author} {\bibfnamefont {I.}~\bibnamefont {Carnimeo}}, \bibinfo {author} {\bibfnamefont {A.}~\bibnamefont {Dal~Corso}}, \bibinfo {author} {\bibfnamefont {S.}~\bibnamefont {de~Gironcoli}}, \bibinfo {author} {\bibfnamefont {P.}~\bibnamefont {Delugas}}, \bibinfo {author} {\bibfnamefont
  {R.~A.}\ \bibnamefont {DiStasio}}, \bibinfo {author} {\bibfnamefont {A.}~\bibnamefont {Ferretti}}, \bibinfo {author} {\bibfnamefont {A.}~\bibnamefont {Floris}}, \bibinfo {author} {\bibfnamefont {G.}~\bibnamefont {Fratesi}}, \bibinfo {author} {\bibfnamefont {G.}~\bibnamefont {Fugallo}}, \bibinfo {author} {\bibfnamefont {R.}~\bibnamefont {Gebauer}}, \bibinfo {author} {\bibfnamefont {U.}~\bibnamefont {Gerstmann}}, \bibinfo {author} {\bibfnamefont {F.}~\bibnamefont {Giustino}}, \bibinfo {author} {\bibfnamefont {T.}~\bibnamefont {Gorni}}, \bibinfo {author} {\bibfnamefont {J.}~\bibnamefont {Jia}}, \bibinfo {author} {\bibfnamefont {M.}~\bibnamefont {Kawamura}}, \bibinfo {author} {\bibfnamefont {H.-Y.}\ \bibnamefont {Ko}}, \bibinfo {author} {\bibfnamefont {A.}~\bibnamefont {Kokalj}}, \bibinfo {author} {\bibfnamefont {E.}~\bibnamefont {Küçükbenli}}, \bibinfo {author} {\bibfnamefont {M.}~\bibnamefont {Lazzeri}}, \bibinfo {author} {\bibfnamefont {M.}~\bibnamefont {Marsili}}, \bibinfo {author} {\bibfnamefont
  {N.}~\bibnamefont {Marzari}}, \bibinfo {author} {\bibfnamefont {F.}~\bibnamefont {Mauri}}, \bibinfo {author} {\bibfnamefont {N.~L.}\ \bibnamefont {Nguyen}}, \bibinfo {author} {\bibfnamefont {H.-V.}\ \bibnamefont {Nguyen}}, \bibinfo {author} {\bibfnamefont {A.}~\bibnamefont {Otero-de-la Roza}}, \bibinfo {author} {\bibfnamefont {L.}~\bibnamefont {Paulatto}}, \bibinfo {author} {\bibfnamefont {S.}~\bibnamefont {Poncé}}, \bibinfo {author} {\bibfnamefont {D.}~\bibnamefont {Rocca}}, \bibinfo {author} {\bibfnamefont {R.}~\bibnamefont {Sabatini}}, \bibinfo {author} {\bibfnamefont {B.}~\bibnamefont {Santra}}, \bibinfo {author} {\bibfnamefont {M.}~\bibnamefont {Schlipf}}, \bibinfo {author} {\bibfnamefont {A.~P.}\ \bibnamefont {Seitsonen}}, \bibinfo {author} {\bibfnamefont {A.}~\bibnamefont {Smogunov}}, \bibinfo {author} {\bibfnamefont {I.}~\bibnamefont {Timrov}}, \bibinfo {author} {\bibfnamefont {T.}~\bibnamefont {Thonhauser}}, \bibinfo {author} {\bibfnamefont {P.}~\bibnamefont {Umari}}, \bibinfo {author}
  {\bibfnamefont {N.}~\bibnamefont {Vast}}, \bibinfo {author} {\bibfnamefont {X.}~\bibnamefont {Wu}},\ and\ \bibinfo {author} {\bibfnamefont {S.}~\bibnamefont {Baroni}},\ }\bibfield  {title} {\bibinfo {title} {Advanced capabilities for materials modelling with quantum espresso},\ }\href {https://doi.org/10.1088/1361-648X/aa8f79} {\bibfield  {journal} {\bibinfo  {journal} {Journal of Physics: Condensed Matter}\ }\textbf {\bibinfo {volume} {29}},\ \bibinfo {pages} {465901} (\bibinfo {year} {2017})}\BibitemShut {NoStop}%
\bibitem [{\citenamefont {Sun}\ \emph {et~al.}(2015)\citenamefont {Sun}, \citenamefont {Ruzsinszky},\ and\ \citenamefont {Perdew}}]{SCAN_PhysRevLett}%
  \BibitemOpen
  \bibfield  {author} {\bibinfo {author} {\bibfnamefont {J.}~\bibnamefont {Sun}}, \bibinfo {author} {\bibfnamefont {A.}~\bibnamefont {Ruzsinszky}},\ and\ \bibinfo {author} {\bibfnamefont {J.~P.}\ \bibnamefont {Perdew}},\ }\bibfield  {title} {\bibinfo {title} {Strongly constrained and appropriately normed semilocal density functional},\ }\href {https://doi.org/10.1103/PhysRevLett.115.036402} {\bibfield  {journal} {\bibinfo  {journal} {Phys. Rev. Lett.}\ }\textbf {\bibinfo {volume} {115}},\ \bibinfo {pages} {036402} (\bibinfo {year} {2015})}\BibitemShut {NoStop}%
\bibitem [{\citenamefont {Zhang}\ \emph {et~al.}(2020)\citenamefont {Zhang}, \citenamefont {Wang}, \citenamefont {Chen}, \citenamefont {Zeng}, \citenamefont {Zhang}, \citenamefont {Wang},\ and\ \citenamefont {E}}]{DPGEN}%
  \BibitemOpen
  \bibfield  {author} {\bibinfo {author} {\bibfnamefont {Y.}~\bibnamefont {Zhang}}, \bibinfo {author} {\bibfnamefont {H.}~\bibnamefont {Wang}}, \bibinfo {author} {\bibfnamefont {W.}~\bibnamefont {Chen}}, \bibinfo {author} {\bibfnamefont {J.}~\bibnamefont {Zeng}}, \bibinfo {author} {\bibfnamefont {L.}~\bibnamefont {Zhang}}, \bibinfo {author} {\bibfnamefont {H.}~\bibnamefont {Wang}},\ and\ \bibinfo {author} {\bibfnamefont {W.}~\bibnamefont {E}},\ }\bibfield  {title} {\bibinfo {title} {Dp-gen: A concurrent learning platform for the generation of reliable deep learning based potential energy models},\ }\href {https://doi.org/https://doi.org/10.1016/j.cpc.2020.107206} {\bibfield  {journal} {\bibinfo  {journal} {Computer Physics Communications}\ }\textbf {\bibinfo {volume} {253}},\ \bibinfo {pages} {107206} (\bibinfo {year} {2020})}\BibitemShut {NoStop}%
\bibitem [{\citenamefont {Thompson}\ \emph {et~al.}(2022)\citenamefont {Thompson}, \citenamefont {Aktulga}, \citenamefont {Berger}, \citenamefont {Bolintineanu}, \citenamefont {Brown}, \citenamefont {Crozier}, \citenamefont {{in 't Veld}}, \citenamefont {Kohlmeyer}, \citenamefont {Moore}, \citenamefont {Nguyen}, \citenamefont {Shan}, \citenamefont {Stevens}, \citenamefont {Tranchida}, \citenamefont {Trott},\ and\ \citenamefont {Plimpton}}]{LAMMPS}%
  \BibitemOpen
  \bibfield  {author} {\bibinfo {author} {\bibfnamefont {A.~P.}\ \bibnamefont {Thompson}}, \bibinfo {author} {\bibfnamefont {H.~M.}\ \bibnamefont {Aktulga}}, \bibinfo {author} {\bibfnamefont {R.}~\bibnamefont {Berger}}, \bibinfo {author} {\bibfnamefont {D.~S.}\ \bibnamefont {Bolintineanu}}, \bibinfo {author} {\bibfnamefont {W.~M.}\ \bibnamefont {Brown}}, \bibinfo {author} {\bibfnamefont {P.~S.}\ \bibnamefont {Crozier}}, \bibinfo {author} {\bibfnamefont {P.~J.}\ \bibnamefont {{in 't Veld}}}, \bibinfo {author} {\bibfnamefont {A.}~\bibnamefont {Kohlmeyer}}, \bibinfo {author} {\bibfnamefont {S.~G.}\ \bibnamefont {Moore}}, \bibinfo {author} {\bibfnamefont {T.~D.}\ \bibnamefont {Nguyen}}, \bibinfo {author} {\bibfnamefont {R.}~\bibnamefont {Shan}}, \bibinfo {author} {\bibfnamefont {M.~J.}\ \bibnamefont {Stevens}}, \bibinfo {author} {\bibfnamefont {J.}~\bibnamefont {Tranchida}}, \bibinfo {author} {\bibfnamefont {C.}~\bibnamefont {Trott}},\ and\ \bibinfo {author} {\bibfnamefont {S.~J.}\ \bibnamefont {Plimpton}},\
  }\bibfield  {title} {\bibinfo {title} {Lammps - a flexible simulation tool for particle-based materials modeling at the atomic, meso, and continuum scales},\ }\href {https://doi.org/https://doi.org/10.1016/j.cpc.2021.108171} {\bibfield  {journal} {\bibinfo  {journal} {Computer Physics Communications}\ }\textbf {\bibinfo {volume} {271}},\ \bibinfo {pages} {108171} (\bibinfo {year} {2022})}\BibitemShut {NoStop}%
\bibitem [{\citenamefont {Chen}\ \emph {et~al.}(2017)\citenamefont {Chen}, \citenamefont {Ko}, \citenamefont {Remsing}, \citenamefont {Andrade}, \citenamefont {Santra}, \citenamefont {Sun}, \citenamefont {Selloni}, \citenamefont {Car}, \citenamefont {Klein}, \citenamefont {Perdew},\ and\ \citenamefont {Wu}}]{333K_pnas.1712499114}%
  \BibitemOpen
  \bibfield  {author} {\bibinfo {author} {\bibfnamefont {M.}~\bibnamefont {Chen}}, \bibinfo {author} {\bibfnamefont {H.-Y.}\ \bibnamefont {Ko}}, \bibinfo {author} {\bibfnamefont {R.~C.}\ \bibnamefont {Remsing}}, \bibinfo {author} {\bibfnamefont {M.~F.~C.}\ \bibnamefont {Andrade}}, \bibinfo {author} {\bibfnamefont {B.}~\bibnamefont {Santra}}, \bibinfo {author} {\bibfnamefont {Z.}~\bibnamefont {Sun}}, \bibinfo {author} {\bibfnamefont {A.}~\bibnamefont {Selloni}}, \bibinfo {author} {\bibfnamefont {R.}~\bibnamefont {Car}}, \bibinfo {author} {\bibfnamefont {M.~L.}\ \bibnamefont {Klein}}, \bibinfo {author} {\bibfnamefont {J.~P.}\ \bibnamefont {Perdew}},\ and\ \bibinfo {author} {\bibfnamefont {X.}~\bibnamefont {Wu}},\ }\bibfield  {title} {\bibinfo {title} {Ab initio theory and modeling of water},\ }\href {https://doi.org/10.1073/pnas.1712499114} {\bibfield  {journal} {\bibinfo  {journal} {Proceedings of the National Academy of Sciences}\ }\textbf {\bibinfo {volume} {114}},\ \bibinfo {pages} {10846} (\bibinfo {year}
  {2017})},\ \Eprint {https://arxiv.org/abs/https://www.pnas.org/doi/pdf/10.1073/pnas.1712499114} {https://www.pnas.org/doi/pdf/10.1073/pnas.1712499114} \BibitemShut {NoStop}%
\bibitem [{\citenamefont {Morrone}\ and\ \citenamefont {Car}(2008)}]{333K_PhysRevLett.101.017801}%
  \BibitemOpen
  \bibfield  {author} {\bibinfo {author} {\bibfnamefont {J.~A.}\ \bibnamefont {Morrone}}\ and\ \bibinfo {author} {\bibfnamefont {R.}~\bibnamefont {Car}},\ }\bibfield  {title} {\bibinfo {title} {Nuclear quantum effects in water},\ }\href {https://doi.org/10.1103/PhysRevLett.101.017801} {\bibfield  {journal} {\bibinfo  {journal} {Phys. Rev. Lett.}\ }\textbf {\bibinfo {volume} {101}},\ \bibinfo {pages} {017801} (\bibinfo {year} {2008})}\BibitemShut {NoStop}%
\bibitem [{\citenamefont {Choi}\ \emph {et~al.}(2017)\citenamefont {Choi}, \citenamefont {Choi}, \citenamefont {Jeon},\ and\ \citenamefont {Cho}}]{Choi_ionagg}%
  \BibitemOpen
  \bibfield  {author} {\bibinfo {author} {\bibfnamefont {J.-H.}\ \bibnamefont {Choi}}, \bibinfo {author} {\bibfnamefont {H.~R.}\ \bibnamefont {Choi}}, \bibinfo {author} {\bibfnamefont {J.}~\bibnamefont {Jeon}},\ and\ \bibinfo {author} {\bibfnamefont {M.}~\bibnamefont {Cho}},\ }\bibfield  {title} {\bibinfo {title} {Ion aggregation in high salt solutions. vii. the effect of cations on the structures of ion aggregates and water hydrogen-bonding network},\ }\href {https://doi.org/10.1063/1.4993479} {\bibfield  {journal} {\bibinfo  {journal} {The Journal of Chemical Physics}\ }\textbf {\bibinfo {volume} {147}},\ \bibinfo {pages} {154107} (\bibinfo {year} {2017})}\BibitemShut {NoStop}%
\bibitem [{\citenamefont {Stauffer}\ and\ \citenamefont {Aharony}(2018)}]{perc2018introduction}%
  \BibitemOpen
  \bibfield  {author} {\bibinfo {author} {\bibfnamefont {D.}~\bibnamefont {Stauffer}}\ and\ \bibinfo {author} {\bibfnamefont {A.}~\bibnamefont {Aharony}},\ }\href@noop {} {\emph {\bibinfo {title} {Introduction to percolation theory}}}\ (\bibinfo  {publisher} {Taylor \& Francis},\ \bibinfo {year} {2018})\BibitemShut {NoStop}%
\bibitem [{\citenamefont {Xu}\ \emph {et~al.}(2013)\citenamefont {Xu}, \citenamefont {Wang}, \citenamefont {Lv},\ and\ \citenamefont {Deng}}]{perc3D}%
  \BibitemOpen
  \bibfield  {author} {\bibinfo {author} {\bibfnamefont {X.}~\bibnamefont {Xu}}, \bibinfo {author} {\bibfnamefont {J.}~\bibnamefont {Wang}}, \bibinfo {author} {\bibfnamefont {J.-P.}\ \bibnamefont {Lv}},\ and\ \bibinfo {author} {\bibfnamefont {Y.}~\bibnamefont {Deng}},\ }\bibfield  {title} {\bibinfo {title} {Simultaneous analysis of three-dimensional percolation models},\ }\href {https://doi.org/10.1007/s11467-013-0403-z} {\bibfield  {journal} {\bibinfo  {journal} {Frontiers of Physics}\ }\textbf {\bibinfo {volume} {9}},\ \bibinfo {pages} {113–119} (\bibinfo {year} {2013})}\BibitemShut {NoStop}%
\end{thebibliography}
\providecommand{\noopsort}[1]{}\providecommand{\singleletter}[1]{#1}%

 \end{document}